# Synthetic Antiferromagnetic Spintronics

R. A. Duine,[1,2] Kyung-Jin Lee,[3,4] Stuart S. P. Parkin,[5,6,7] and M. D. Stiles[8]

[1)]*Institute for Theoretical Physics, Universiteit Utrecht, Leuvenlaan 4, 3584 CE Utrecht,*

*The Netherlands*

[2)]*Department of Applied Physics, Eindhoven University of Technology, P.O. Box 513, 5600 MB Eindhoven,*

*The Netherlands*

[3)]*Department of Materials Science and Engineering, Korea University, Seoul 02841, Korea*

[4)]*KU-KIST Graduate School of Converging Science and Technology, Korea University, Seoul 02841, Korea*

[5)] *Max Planck Institute for Microstructure Physics, Halle (Saale) D-06120, Germany.*

[6)] *IBM Research–Almaden, San Jose, CA 95120, USA.*

[7)]*Institute of Physics, Martin Luther University Halle-Wittenberg, Halle (Saale) D-06120, Germany*

[8)]*Center for Nanoscale Science and Technology, National Institute of Standards and Technology,*

*Gaithersburg, Maryland 20899, USA*

(Dated: 20 May 2017)

Spintronic and nanomagnetic devices often derive their functionality from layers of different materials and the interfaces between them. This is especially true for synthetic antiferromagnets — two or more ferromagnetic layers that are separated by metallic spacers or tunnel barriers and which have antiparallel magnetizations. Here, we discuss the new opportunities that arise from synthetic antiferromagnets, as compared to crystal antiferromagnets or ferromagnets.

# Introduction

Advances in nanofabrication techniques for magnetic materials — such as Fe, Ni, Co, Cr and their alloys — have, since the late 1980's, enabled researchers to engineer stacks of thin (nanometers) layers of magnetic and nonmagnetic material. The study of such magnetic multilayers and superlattices – i.e., periodic multilayers – has led to many discoveries and potential applications. The first among these is the existence of a coupling between two magnetic layers adjacent to the same non-magnetic spacer.[1–3] This interlayer exchange coupling is essentially a spin-dependent Ruderman-Kittel-Kasuya-Yosida (RKKY) coupling that is rooted in Friedel-like spatial oscillations in the spin density of the non-magnetic spacer that are caused by the adjacent ferromagnets. The oscillating spin density in turn leads to an interlayer exchange coupling constant that oscillates with the distance between the ferromagnetic layers.[4–8] By changing the thickness of non-magnetic material between two magnetic layers one can therefore tune the interaction from ferromagnetic — preferring parallel alignment — to antiferromagnetic, preferring antiparallel alignment, whereas for thick spacers the interlayer exchange coupling is suppressed. Trilayers, multilayers or superlattices, in which the interaction between magnetic layers is antiferromagnetic are now commonly referred to as synthetic antiferromagnets (see Figure). The antiferromagnetic coupling was crucial for the discovery that the resistance of metallic magnetic multilayers depends on the relative orientation of the magnetization in adjacent layers.[9,10] This finding — called giant magnetoresistance (GMR) or, in the case of a tunneling barrier, tunneling magnetoresistance (TMR) — kickstarted the field of nanomagnetism and spintronics.

Before we discuss the physics of GMR in somewhat more detail, let us first compare synthetic with crystal antiferromagnets, i.e., the antiferromagnets found in nature as bulk single crystals. Most important is the difference in physical origin between the interlayer exchange coupling described above and the direct exchange or superexchange coupling in crystal antiferromagnets. As a result, the interlayer exchange coupling is much weaker and typically two or more orders of magnitude smaller than ordinary exchange coupling. This relatively weak exchange coupling implies that external fields and anisotropy can compete with the antiferromagnetic order in synthetic antiferromagnets, but in crystal antiferromagnets an external field can typically only reorient the magnetization, for example near a spin-flop transition. For a synthetic antiferromagnet, a transition from antiparallel to parallel arrangement of the magnetization direction in the magnetic layers can therefore be achieved by magnetic fields that can be easily reached in a laboratory. This was an essential part of the discovery of GMR.

Furthermore, the repeat distance of the antiferromagnetic order in synthetic antiferromagnets is larger than in crystal antiferromagnets. While in the latter the magnetic order alternates on atomic length scales, the layer thickness in magnetic multilayers is typically several nanometers. For most situations the electron motion within one magnetic layer is therefore appropriately described by a spin-dependent semi-classical model.[11] GMR, for instance, is typically modelled by taking into account electron diffusion within the magnetic layers supplemented with spin-dependent resistances of the various layers and interfaces between them, as well as spin relaxation. For crystal antiferromagnets this picture breaks down as the electrons are in that case phase coherent over a region that is larger than the length scale of the antiferromagnetic order.

The large tunability of synthetic antiferromagnets via layer thickness and material composition, together with the above-mentioned unique energy and spatial scales, leads to physical phenomena that are either very different from, or have no counterpart in, crystal antiferromagnets. In the remainder of this commentary we review some of these phenomena and discuss possible new directions.

**Statics and dynamics**

One of the most basic applications of the tunability of magnetic multilayers is in magnetic field sensing.[12] At a rudimentary level, field sensors consist of a layer stack that exhibits GMR or TMR and that typically incorporates a free layer and a pinned layer. Interlayer exchange coupling and the remaining dipolar fields cause an offset of the response curve of the free layer — the resistance as a function of field is no longer symmetric around zero field. Large offset fields degrade the performance of sensors that operate in the linear regime by reducing the change of resistance with changes in field. To minimize dipolar fields, and thereby reduce hysteresis and increase sensitivity, the pinned is chosen to be a synthetic antiferromagnet. Recent results demonstrate that the tunability of magnetic multilayers allows for simultaneous optimization of dipolar and offset fields.[13–15]

The relative orientation between pinned and free layers is used to encode binary information in magnetic random access memory (MRAM).[16] Two key physical properties for the design of such memories are the thermal stability, determined by the energy barrier that separates the two states of one bit, and the current or field required to switch, i.e., write the bit. For the goal of keeping thermal stability high while reducing the energy required to write the bit, current-induced switching that uses spin-polarized currents is most attractive.[17] It was shown that using synthetic antiferromagnets can reduce the critical current for switching while maintaining thermal stability.[18] Moreover, it was predicted that, owing to the dynamics resulting from interlayer exchange coupling within a synthetic antiferromagnet free layer, the time for switching can be significantly reduced.[19]

En route to demonstrating switching, spin-current-driven auto-oscillations in devices with a synthetic antiferromagnetic layer were demonstrated.[20] These are characterized by high power of emission, narrow linewidth, and nontrivial field and current dependence, due to the multiple layers whose dynamics are governed by the interlayer exchange.[21,22] Linewidth broadening of dynamic modes in synthetic antiferromagnetic superlattices, on the other hand, has been reported in Ref. 22 and was interpreted to result from spin pumping from the magnetic layers into the nonmagnetic spacers. Switching by spin-orbit torques has recently been demonstrated in devices containing synthetic antiferromagnetic layers.[23,24] Such switching requires a symmetry breaking field. In Ref. 23 this was provided by the interlayer exchange coupling between an in-plane and out-of-plane magnetized layer, while in Ref. 25 exchange bias was used.

**Domain walls and solitons**

The motion of domain walls and solitons in synthetic antiferromagnets has been studied both theoretically and experimentally.[26–32] Due to partial cancellation of dipolar fields, the domain walls in synthetic antiferromagnets tend to be narrower than in ferromagnetic wires. The narrow walls are beneficial for the design of race track memories based on driving trains of closely-spaced domain walls. Saariskoski *et al.*[27] showed that the interlayer exchange coupling has two important novel features for domain wall motion driven by spin transfer torques. First, the interlayer exchange coupling leads to an attraction between the domain walls in the two ferromagnetic layers that make up the synthetic antiferromagnet. This coupling reduces effects of pinning as both domain walls mutually assist each other in overcoming potential barriers. Second, the attraction between domain walls in different ferromagnetic layers of the synthetic antiferromagnet leads to increased velocities of the domain walls if they are not on top of each other. While domain walls in crystal antiferromagnets are also expected to move at large – in comparison to ferromagnets – velocities, the underlying physics is different because of their much stronger antiferromagnetic exchange and the much shorter length scales. The narrow walls and fast domain motion have been seen experimentally in a reduced threshold current for domain wall motion in a synthetic ferrimagnet in Ref. 33. Indirect support was obtained via noise measurements in synthetic antiferromagnets, while Ref. 24 attributed switching behaviour in a multilayer to domain wall motion and nucleation.

Most of the ongoing research on current-driven domain wall motion focuses on multilayers that involve heavy elements with strong spin-orbit coupling, such as Pt or Ta, as the nonmagnetic layers. This spin-orbit coupling has several new physical consequences. First of all, the boundary between heavy nonmagnetic and magnetic metal leads to interface-induced Dzyaloshinskii-Moriya (DM) interactions (see box). These interactions lead to chiral domain walls — domain walls in which the spins have a preferred sense of rotation. In particular, the interfacial DM interactions stabilize Néel domain walls that are efficiently driven by spin-orbit torques. These spin-orbit torques are also induced by the spin-orbit coupling in the heavy metallic layer. The interlayer exchange coupling stabilizes the Néel structure of the walls such that they can be driven more efficiently by spin-orbit torques. On top of this, the interlayer exchange coupling leads to additional torques that efficiently drive the domain walls in both ferromagnetic layers of the synthetic antiferromagnet in the same direction. It was experimentally shown in Ref. 28 that large domain wall velocities (of up to 750 m/s) are obtained for domain walls in synthetic antiferromagnets of Co/Ni magnetic layers, separated by thin layers of Ru.

A completely different type of domain wall motion was demonstrated by Lavrijsen *et al.*[31]. The domain walls considered in this work are kink defects in the antiferromagnetic order of the synthetic antiferromagnet such that two adjacent magnetic layers have magnetizations that are parallel, rather than antiparallel. In a superlattice designed with different interlayer exchange coupling and different magnetic layer thicknesses, these authors were able to demonstrate injection and propagation of kinks by external field pulses. This latter work is an attractive example of how the large tunability of synthetic antierromagnets, and magnetic multilayers in general, can be put to use to enable new functionalities.

**Outlook**

One of the areas of interest in synthetic antiferromagnets is magnetic skyrmions. While there has not been much work yet in this direction, Ref. 35 pointed out that the Magnus force that acts on a skyrmion — or more generally on two-dimensional magnetic structures that have a nonzero winding number — and pushes them sideways is counteracted and cancelled by the interlayer exchange coupling in synthetic antiferromagnets. This is beneficial for applications in which skyrmions are driven along narrow wires, as the sideways motion may cause the skyrmions to interact with the edges of the wire and disappear. Another attractive research direction is to alter the magnetic properties of the synthetic antiferromagnets *in-situ* by electric fields,[36] rather than by engineering different systems with different properties. For example, it was theoretically proposed in Ref. 37 that the interlayer exchange coupling can be switched from ferromagnetic to antiferromagnetic either by an electric field or by making use of a ferroelectric layer. These and other examples ultimately show that synthetic antiferromagnets can in some sense be thought of as materials with properties in between those of ferromagnets and antiferromagnets. Some of their properties derive from the ferromagnetic layers that constitute them, whereas other properties derive from the coupling between these layers. Engineering and exploiting their tunability will surely lead to new physics and applications for the years to come.

**Acknowledgments**

RD is supported by the Stichting voor Fundamenteel Onderzoek der Materie (FOM), the European Research Council (ERC), and is part of the D-ITP consortium, a program of the Netherlands Organization for Scientific Research (NWO) that is funded by the Dutch Ministry of Education, Culture and Science. K.-J.L. was supported by the National Research Foundation of Korea (NRF) (NRF-2015M3D1A1070465, NRF-2017R1A2B2006119).

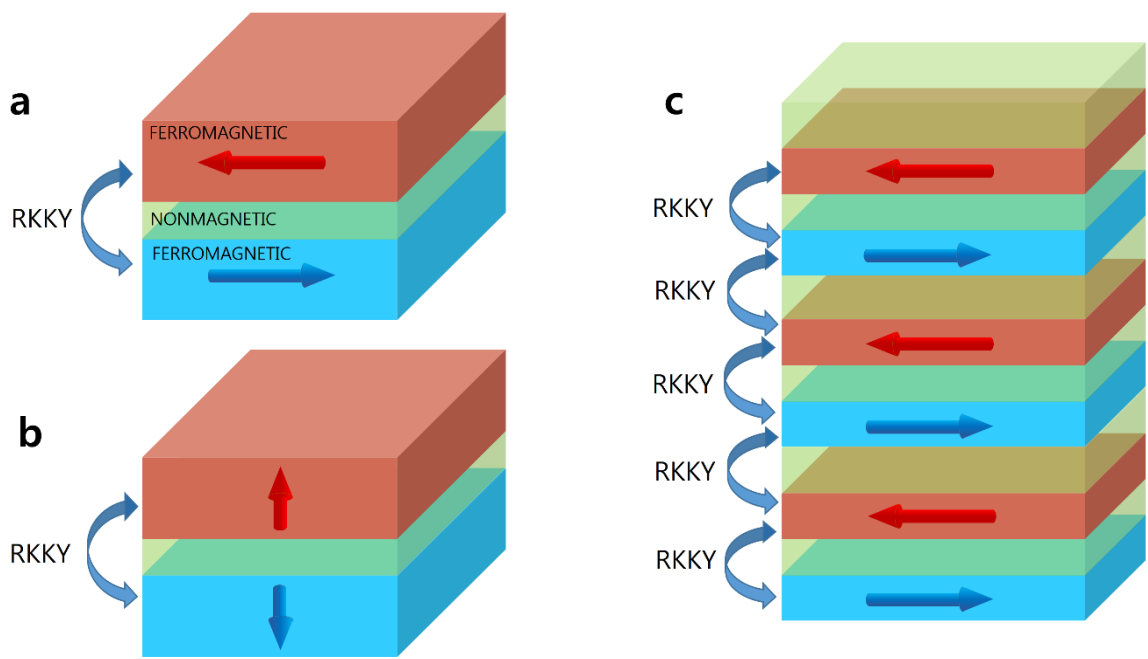

Figure: Schematic of synthetic antiferromagnets. a, bilayers with in-plane magnetization. b, bilayers with out-of-plane magnetizations. c, multilayers. The arrows within each ferromagnetic layer indicate the direction of magnetization.

**BOX – DZYALOSHINSKII-MORIYA (DM) EXCHANGE INTERACTIONS**

Apart from the well-known Heisenberg-type exchange interactions between spins in ferromagnetic materials, there can exist so-called Dzyaloshinskii-Moriya (DM) interactions in magnetic systems that lack a center of inversion and exhibit spin-orbit coupling. In the situation of two magnetic atoms (spin $S_1$ and $S_2$) in presence of a third non-magnetic atoms with spin-orbit coupling (see figure) this interaction has the form $\sim \mathbf{D} \cdot \mathbf{S}_1 \times \mathbf{S}_2$, with $\mathbf{D}$ the Dzyaloshinskii vector and $\mathbf{R}_1$ and $\mathbf{R}_2$ the respective positions of the magnetic atoms with respect to the non-magnetic one.

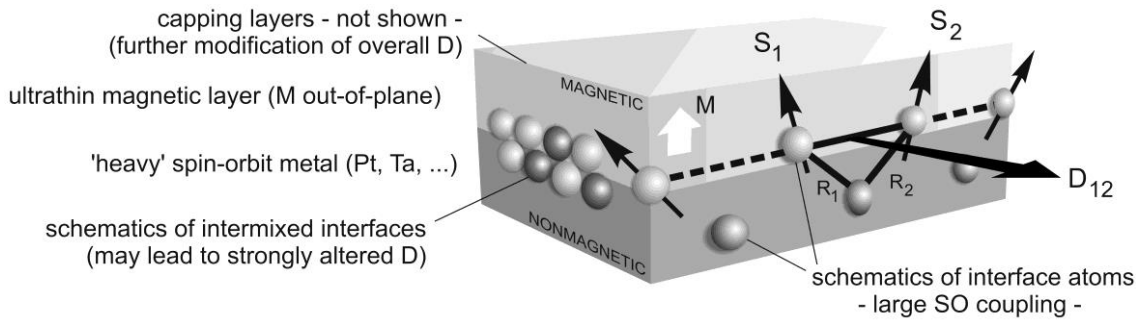

This interaction clearly favors a certain misaligned and turning sense (chirality) of the magnetic moments. Most important for magnetic multilayers are the DM interactions induced by interfaces between magnetic metals and metals with strong spin-orbit coupling. Figure adapted from Nature Nanotech. 8, 152 (2013).